\documentstyle[pre,aps,preprint,eqsecnum,epsf]{revtex}

\tightenlines

\begin{document}

\title{Chaotic Symmetry Breaking and Dissipative Two-Field Dynamics}

\author{Rudnei O. Ramos\thanks{Email address: rudnei@dft.if.uerj.br}
and F. A. R. Navarro}

\address{{\it Departamento de F\'{\i}sica Te\'orica, 
Instituto de F\'{\i}sica, Universidade do Estado do Rio de Janeiro,\\
20550-013 Rio de Janeiro, RJ, Brazil}}

\maketitle

\begin{abstract}

The dynamical symmetry breaking in a two-field model is studied by
numerically solving the coupled effective field equations. These are
dissipative equations of motion that can exhibit strong chaotic
dynamics. By choosing very general model parameters leading to symmetry
breaking along one of the field directions, the symmetry broken vacua
make the role of transitory strange attractors and the field
trajectories in phase space are strongly chaotic. Chaos is quantified by
means of the determination of the fractal dimension, which gives an
invariant measure for chaotic behavior. Discussions concerning chaos and
dissipation in the model and possible applications to related problems
are given.

\vspace{0.34cm}
\noindent
PACS number(s): 11.10.Wx, 05.45.Df, 98.80.Cq

\end{abstract}

\newpage

\section{Introduction}

The study and understanding of the dynamics of fields are a timely
subject and of broad interest, with applications in diverse areas like
in particle physics, cosmology and in condensed matter (for a recent
review, see \cite{boya1} and references therein). Additional interest on
the subject comes from the fact that many of the theoretical ideas and
models can be tested in ongoing experiments, as those been performed in
condensed matter systems, and in the future ones, in the entering of
operation of the RHIC and LHC heavy ion colliders, which will be able to
probe possible new phenomena at the QCD scale and space based
experiments, which will be putting on test different cosmological
models. It is then becoming urgent the detailed investigation of the
underline field dynamics that may be common to all these very different
areas of physical research.

In this paper we are mainly concerned with the connection between the
development of strong nonlinearities in the time evolving system of
equations of motion of a given field theory model and the possible
chaotic behavior associated to them. Lets recall that in symmetry
breaking phase transitions we are usually interested in the study of the
evolution of a given order parameter, for example the magnetization in
spin systems in statistical physics or a vacuum expectation value of
some scalar field (the Higgs) in particle physics, which gives a measure
of the degree of organization of the system at the macroscopic level.
However, at the microscopic level disorder is related to the
nonlinearities and fluctuations responsible to chaotic behavior in the
system and these chaotic motion phenomena can reflect in a nontrivial
time dependence of the macroscopic quantities and, therefore,
influencing all the dynamics of the system. This is clear once several
properties of the system at longer times are closely related to the
microscopic physics, like relaxation to equilibrium, phase ordering,
thermalization and so on. Thus we expect that chaos will be not only an
important ingredient in determining the final states of a given system
but also in how it gets there.

Previous studies on chaos in field theory have mostly emphasized chaotic
behavior in gauge theory models (for a review and additional references
and applications, see \cite{biro}). In special, in homogeneous
Yang-Mills-Higgs models we can reduce the system of classical equations
of motion to ones analogous to those of nonlinear coupled oscillators,
which is well known to exhibit chaotic motion. 

While in all the previous works on chaotic dynamics of fields dealt with
(conservative) Hamiltonian systems, here we will be mainly concerned
with the {\it effective } field evolution equations, which are known to
be intrinsically dissipative \cite{GR,boya,greiner,BGR,BGR2,hu,ian} and,
therefore, the dynamical system we will be studying is non-Hamiltonian.

Thus, we will be concerned with the influence of field dissipation, due
to field decaying modes, on the degree of chaoticity of the field
dynamics. Typically we expect that dissipation damps the fluctuations on
the system and consequently tends to suppress possible chaotic motions
and makes field trajectories in phase space to tend faster to the system
asymptotic states. On the other hand there are well known examples of
dynamical systems, as the Lorenz system \cite{ott}, which are
dissipative ones and at the same time they display a very rich evolution
on phase space in which, under appropriate system parameters, field
trajectories may be lead towards strange attractors. The verification of
the same properties in a model motivated by particle physics would be
a novel result with possible consequences to, for example, particle
physics phenomenology and cosmology.

We study chaos in our dynamical system of equations by means of the
measure of the fractal dimension (or dimension information) [for a
review and definitions, see {\it e.g.}, \cite{ott}], which gives a
topological measure of chaos for different space-time settings and it is
a quantity invariant under coordinate transformations, providing then an
unambiguous signal for chaos \cite{levin,levin2}. The method we apply in this
work for quantifying chaos will then be particularly useful in our
planned future applications of our model and it extensions to a
cosmological context, in which case other methods may be ambiguous,
like, for example, the determination of Lyapunov exponents, which
does not give a coordinate invariant measure for chaos, as discussed in
\cite{levin,levin2}. Also, other methods for studying chaotic systems, like for
example by Poincar\'e sections, are not suitable in the case we are
interested here, in which chaos is a transitory phenomenon as we will
see.

This work is organized as follows. In Sec. II we introduce the model and
discuss its general properties at the classical level. In Sec. III
we obtain the one-loop effective equations of motion for the fields
and we determine the general form of the
nonlocal (non-Markovian) dissipative kernels.
In Sec. IV we then
discuss the validity of the one-loop approximation and the
Markovian approximation for the dissipative kernels appearing in
the one-loop effective equations. We couple our system of fields
to a set of $N$ other fields making up the bath (or environment),
in which the system evolves, and the large N behavior of the various
important quantities is determined. In Sec. V we then present the 
Markovian form for the effective equations of motion and our main
numerical results, where we determine the fractal dimension. In
Sec. VI our concluding remarks are given. 

\section{The Two Interacting  Scalar Fields Model}

The model we will study consists of two scalar fields in
interaction with Lagrangian density given by

\begin{eqnarray}
{\cal L} [\Phi, \Psi] &=& \frac{1}{2} (\partial_\mu \Phi)^2 -
\frac{m_\phi^2}{2} \Phi^2 - \frac{\lambda_\phi}{4 !} \Phi^4
\nonumber \\
& + &
\frac{1}{2} (\partial_\mu \Psi)^2 -
\frac{m_\psi^2}{2} \Psi^2 - \frac{\lambda_\psi}{4 !} \Psi^4
-\frac{g^2}{2}\Phi^2 \Psi^2 \;.
\label{Lpp}
\end{eqnarray}

\noindent 
All coupling constants are positive and $m_\phi^2 > 0$, but we
choose $m_\psi^2 < 0$, such that we allow for spontaneous symmetry
breaking in the $\Psi$-field direction. Additionally, note that from the
above Lagrangian, that for values of $\Phi$ larger than a $\Phi_{\rm cr}$,
where $\Phi^2_{\rm cr} = |m_\psi^2|/g^2$, there is no symmetry breaking
in the $\Psi$-field direction. Thus, for example, if we have an initial
state prepared at $\Phi > \Phi_{\rm cr}$, the $\Psi$ field will move
towards zero and remain around that state till eventually $\Phi$ crosses
below the critical value inducing a (dynamical) symmetry breaking in the
$\Psi$-field direction, after which the fields evolve to their vacuum
values at $\langle \Psi \rangle_v = \pm \sqrt{6}
|m_\psi|/\sqrt{\lambda_\psi}$ and $\langle \Phi \rangle_v =0$. 

The classical equations of motion for field configurations
$\langle \Phi \rangle = \phi_c$ and $\langle \Psi \rangle =
\psi_c$ can be readily be obtained from
Eq. (\ref{Lpp}):

\begin{equation}
\Box \phi_c + m_\phi^2 \phi_c + \frac{\lambda_\phi}{6}\phi_c^3 +
g^2\phi_c\psi_c^2 = 0  
\label{phiclass}\;,
\end{equation}

\begin{equation}
\Box \psi_c + m_\psi^2 \psi_c + \frac{\lambda_\psi}{6}\psi_c^3 + g^2\psi_c
\phi_c^2 = 0 \;.
\label{psiclass}
\end{equation}

\noindent
{}For homogeneous fields $\phi_c$ and $\psi_c$ 
(in which case Eqs. (\ref{phiclass}) and (\ref{psiclass})
are equivalent to the equations of motion of two particles with quartic
potentials and quadratic interaction between them)
the above equations
are well known to lead to chaotic trajectories in phase space.
The chaotic behavior of very similar classical equations have 
been studied recently in Ref. \cite{bazeia} for a model with
$Z_2 \times Z_2$ symmetry and the corresponding
dynamical system shown to be chaotic for symmetry breaking in one
of the field directions.

\section{The One-Loop Equations of Motion}

In the
following we study how quantum effects, which will be responsible for the
appearance of dissipative dynamics in the system evolution, will
influence this dynamical symmetry breaking process and we identify and
quantify possible chaotic behavior in the system evolution, as a
function of the fields dissipations.

Roughly speaking, chaos means extreme sensitivity to small changes in
the initial conditions. Due to nonlinearity, fluctuations in the initial
conditions of chaotic systems evolve such that they can completely alter
the asymptotic outcome of the unperturbed trajectories in phase space.
We here quantify the chaoticity of the system by measuring the fractal
dimension.
The fractal dimension is associated with the possible different exit
modes under small changes of initial conditions and it gives a measure
of the degree of chaos of a dynamical system \cite{ott}. The exit modes
we refer to above are one of the symmetry breaking minima in the
$\Psi$-field direction, which are attractors of field trajectories in
phase space. The method we employ to determine the fractal dimension is
the box-counting method, whose definition and the specific numerical
implementation we use here have been described in details in Ref.
\cite{fractal}.

Quantum corrections are taken into account in the effective equations of
motions (EOMs) by use of the tadpole method \cite{tadpole}. Let $\phi_c$ and
$\psi_c$ be the expectation values for $\Phi$ and $\Psi$, respectively.
Splitting the fields in (\ref{Lpp}) in the expectation values and
fluctuations, 

\begin{equation}
\Phi \to \phi_c + \phi \;\;\; {\rm and} \;\;\; \Psi \to \psi_c + \psi,
\end{equation}

\noindent
where $\langle \Phi \rangle = \phi_c$ and $\langle \Psi \rangle =
\psi_c$, the EOMs for $\phi_c$ and $\psi_c$ are obtained by imposing
that $\langle \phi \rangle = 0$ and $\langle \psi \rangle = 0$, which
lead to the condition that the sum of all tadpole terms for each field
vanishes. Restricting our analysis of the EOMs to homogeneous fields
($\phi_c \equiv \phi_c (t)$, $\psi_c \equiv \psi_c (t)$), thus, at the
one-loop order we can write the following EOMs for $\phi_c$ and $\psi_c$,
respectively, as\footnote{Note that the mixed field averages in
(\ref{eqphi}) and (\ref{eqpsi}) can only be treated within perturbation
theory.} (where overdots mean time derivatives) 

\begin{eqnarray}
\ddot{\phi}_c &+& m_\phi^2 \phi_c + \frac{\lambda_\phi}{6} \phi_c^3 
+  g^2 
\phi_c \psi_c^2 +
\frac{\lambda_\phi}{2} \phi_c \langle \phi^2 \rangle \nonumber \\
& + &   g^2 
\phi_c \langle \psi^2 \rangle + 2 g^2 \psi_c \langle \phi \psi \rangle = 0 
\label{eqphi}
\end{eqnarray}

\noindent
and
\begin{eqnarray}
\ddot{\psi}_c &+& m_\psi^2 \psi_c + \frac{\lambda_\psi}{6} \psi_c^3 
+  g^2 
\psi_c \phi_c^2 +
\frac{\lambda_\psi}{2} \psi_c \langle \psi^2 \rangle \nonumber \\
& + &    g^2 
\psi_c \langle \phi^2 \rangle + 2 g^2 \phi_c \langle \phi \psi \rangle= 0 \;,
\label{eqpsi}
\end{eqnarray}

\noindent 
where $\langle \phi^2 \rangle$ and $\langle \psi^2 \rangle$
are given in terms of the coincidence limit of the (causal) two-point
Green's functions $G^{++}_\phi (x,x')$ and $G^{++}_{\psi} (x,x')$, which
are obtained from the $(1,1)$-component of the real time matrix of full
propagators which satisfy the appropriate Schwinger-Dyson equations
(see, {\it e.g.}, Refs. \cite{BGR} and \cite{ian} for further details):

\begin{eqnarray}
\lefteqn{\left[\Box + m_\phi^2 + \frac{\lambda_\phi}{2} \phi_c^2 + g^2
\psi_c^2 \right] G_\phi (x,x')}  \nonumber \\
& & + \int d^4 z \Sigma_\phi (x,z) G_\phi (z,x') = i \delta
(x,x') \;, 
\label{Gphi}
\end{eqnarray}

\noindent
and

\begin{eqnarray}
\lefteqn{\left[ \Box + m_\psi^2 + \frac{\lambda_\psi}{2} \psi_c^2 + g^2
\phi_c^2 \right] G_{\psi} (x,x') }  \nonumber \\
& & + \int d^4 z \Sigma_{\psi} (x,z) G_{\psi} (z,x') = i \delta
(x,x')\;,  
\label{Gpsi}
\end{eqnarray}

\noindent
where $\Sigma_{\phi} (x,x')$ and $\Sigma_{\psi} (x,x')$ are the self-energies
for $\Phi$ e $\Psi$, respectively. The momentum-space Fourier transform
of $G (x,x')$ (for both fields) can be expressed in the form

\begin{equation}
G(x,x') = i \int \frac{d^3 q}{(2 \pi)^3} 
e^{i {\bf q} . ({\bf x} - {\bf x}')}
\left(
\begin{array}{ll}
G^{++}({\bf q}, t- t') & \:\: G^{+-}({\bf q}, t-t') \\
G^{-+}({\bf q}, t- t') & \:\: G^{--}({\bf q}, t-t')
\end{array}
\right) \: ,
\label{Gmatrix}
\end{equation}

\noindent
where

\begin{eqnarray}
\lefteqn{G^{++}({\bf q} , t-t') = G^{>}({\bf q},t-t')
\theta(t-t') + G^{<}({\bf q},t-t') \theta(t'-t)}
\nonumber \\
& & G^{--}({\bf q} , t-t') = G^{>}({\bf q},t-t')
\theta(t'-t) + G^{<}({\bf q},t-t') \theta(t-t')
\nonumber \\
& & G^{+-}({\bf q} , t-t') = G^{<}({\bf q},t-t')
\nonumber \\
& & G^{-+}({\bf q},t-t') = G^{>}({\bf q},t-t')\;,
\label{G of k}
\end{eqnarray}

\noindent
and the fully dressed two-point functions, at a finite temperature $T=1/\beta$,
are given by\footnote{R.O.R.
thanks I. Lawrie for pointing him the correct form for these expressions.}

\begin{eqnarray}
G^{>}({\bf q} ,t-t') &= &
\frac{1}{2\omega}\left\{[1+n(\omega-i\Gamma)]
e^{-i(\omega-i\Gamma)(t-t')}
+ n(\omega+i\Gamma)e^{i(\omega+i\Gamma)(t-t')}\right\}\theta(t-t')
\nonumber\\
&+&\frac{1}{2\omega}\left\{[1+n(\omega+i\Gamma)]
e^{-i(\omega+i\Gamma)(t-t')}
+ n(\omega-i\Gamma)e^{i(\omega-i\Gamma)(t-t')}\right\}\theta(t'-t)\;,
\nonumber \\
G^{<}({\bf q} , t-t') & = & G^{>}({\bf q}, t'-t) \: ,
\label{G><}
\end{eqnarray}

\noindent
where $n(\omega)$ is the Bose
distribution, $\omega\equiv \omega({\bf q})$ is the particle's 
dispersion relation and $\Gamma$
is the decay width, defined as usual in terms of the field self-energy by

\begin{equation}
\Gamma (q) =  
\frac{{\rm Im} \Sigma ({\bf q},\omega)}
{2 \omega}\;.
\label{width}
\end{equation}

Assuming the couplings
$g,\lambda_\phi,\lambda_\psi \ll 1$, so that  
perturbation theory
can be consistently formulated and subleading terms can be neglected,
then by perturbatively expanding the field averages in Eqs. (\ref{eqphi}) 
and (\ref{eqpsi}), we can write
for the EOMs

\begin{eqnarray}
\lefteqn{\ddot{\phi}_c(t) + \tilde{m}_\phi^2 \phi_c (t) + 
\frac{\lambda_\phi}{6} 
\phi_c^3 (t) + g^2 
\phi_c(t) \psi_c^2(t)
} \nonumber \\
& & + \lambda_\phi \phi_c (t) \int_{-\infty}^t dt'\left[ 
\frac{\lambda_\phi}{2} \phi_c^2 (t') + g^2 \psi_c^2 (t') 
\right] \int \frac{d^3 {\bf q}}{(2 \pi)^3} 
{\rm Im} \left[ G_\phi^{++} ({\bf q},t-t') \right]^2 \nonumber \\
& & + 2 g^2 \phi_c (t)  \int_{-\infty}^t  dt'  
\left[ \frac{\lambda_\psi}{2} \psi_c^2 (t') + g^2 \phi_c^2 (t') \right]
\int \frac{d^3 {\bf q}}{(2 \pi)^3} 
{\rm Im} \left[ G_\psi^{++} ({\bf q},t-t') \right]^2 \nonumber \\
& & + 8 g^4 \psi_c (t)
\int_{-\infty}^{t} dt' \phi_c (t') \psi_c(t')
\int \frac{d^3 {\bf q}}{(2 \pi)^3} 
{\rm Im} \left[ G_\phi^{++} ({\bf q},t-t') G_\psi^{++} ({\bf q},t-t') 
\right] = 0
\label{eqphi2}
\end{eqnarray}

\noindent
and
\begin{eqnarray}
\lefteqn{\ddot{\psi}_c(t) + \tilde{m}_\psi^2 \psi_c (t) + 
\frac{\lambda_\psi}{6} 
\psi_c^3 (t) + g^2 
\psi_c (t) \phi_c^2 (t)
} \nonumber \\
& & + \lambda_\psi \psi_c (t)  \int_{-\infty}^t dt'
\left[ \frac{\lambda_\psi}{2} \psi_c^2 (t') +
g^2 \phi_c^2 (t') \right] \int \frac{d^3 {\bf q}}{(2 \pi)^3} 
{\rm Im} \left[ G_\psi^{++} ({\bf q},t-t') \right]^2 \nonumber \\
& & + 2 g^2 \psi_c (t)  \int_{-\infty}^t  dt' 
\left[ \frac{\lambda_\phi}{2} \phi_c^2 (t') + g^2 \psi_c^2 (t') \right]
\int \frac{d^3 {\bf q}}{(2 \pi)^3} 
{\rm Im} \left[ G_\phi^{++} ({\bf q},t-t') \right]^2 \nonumber \\ 
& & + 8 g^4 \phi_c (t)
\int_{-\infty}^{t}  dt' \phi_c (t') \psi_c(t')
\int \frac{d^3 {\bf q}}{(2 \pi)^3} 
{\rm Im} \left[ G_\phi^{++} ({\bf q},t-t') G_\psi^{++} ({\bf q},t-t') 
\right] = 0,
\label{eqpsi2}
\end{eqnarray}
 
\noindent 
where

\begin{eqnarray}
\tilde{m}_\phi^2 &=& m_\phi^2 +\frac{\lambda_\phi}{2}
\langle\phi^2\rangle_0 +g^2 \langle\psi^2\rangle_0\;,
\nonumber \\
\tilde{m}_\psi^2 &=& m_\psi^2 +\frac{\lambda_\psi}{2} 
\langle\psi^2\rangle_0 +g^2
\langle\phi^2\rangle_0\;,
\label{masses}
\end{eqnarray}

\noindent 
with $\langle \ldots\rangle_0$ meaning fields
independent averages. The nonlocal terms of the type appearing in the
above equations have been shown in Refs. \cite{GR,boya,greiner,BGR} to
lead to dissipative dynamics in the EOMs. This can be made more explicit
by an appropriate
integration by parts in the time integrals in Eqs. (\ref{eqphi2}) and
(\ref{eqpsi2}) (see Ref. \cite{boya}) to obtain the result

\begin{eqnarray}
\lefteqn{\ddot{\phi}_c(t) + \tilde{m}_\phi^2 \phi_c (t) + 
\frac{\tilde{\lambda}_\phi}{6} 
\phi_c^3 (t) + \tilde{g}^2 
\phi_c(t) \psi_c^2(t)
} \nonumber \\
& & + \phi_c (t) \int_{-\infty}^t dt' 
\phi_c (t') \dot{\phi}_c (t') F_1 (t,t')  
+ \phi_c (t)  \int_{-\infty}^t  dt'  
\psi_c (t') \dot{\psi}_c (t') F_3 (t,t')
\nonumber \\
& & + \psi_c (t)
\int_{-\infty}^{t} dt' \left[\phi_c (t') \dot{\psi}_c (t')+
\dot{\phi}_c (t') \psi_c (t')\right] F_4 (t,t') = 0\;,
\label{eqphi2_1}
\end{eqnarray}

\noindent
and
\begin{eqnarray}
\lefteqn{\ddot{\psi}_c(t) + \tilde{m}_\psi^2 \psi_c (t) + 
\frac{\tilde{\lambda}_\psi}{6} 
\psi_c^3 (t) + \tilde{g}^2 
\psi_c (t) \phi_c^2 (t)
} \nonumber \\
& & + \psi_c (t)  \int_{-\infty}^t dt'
\psi_c (t') \dot{\psi}_c (t') F_2 (t,t') 
+ \psi_c (t)  \int_{-\infty}^t  dt' 
\phi_c (t') \dot{\phi}_c (t') F_3 (t,t') 
\nonumber \\ 
& & + \phi_c (t)
\int_{-\infty}^{t}  dt' 
\left[ \phi_c (t') \dot{\psi}_c (t') + \dot{\phi}_c (t') \psi_c(t')\right]
{}F_4 (t,t') =0\;,
\label{eqpsi2_1}
\end{eqnarray}
 
\noindent 
where the dissipative kernels $F_1, F_2, F_3$ and $F_4$ are given by 

\begin{eqnarray}
{}F_1 (t,t') &=& -\int dt' \int \frac{d^3 {\bf q}}{(2 \pi)^3}
\left\{\lambda_\phi^2
{\rm Im} \left[ G_\phi^{++} ({\bf q},t-t') \right]^2 
+ 4 g^4  {\rm Im} \left[ G_\psi^{++} ({\bf q},t-t') \right]^2 
\right\}\;,
\nonumber \\
{}F_2 (t,t') &=&-\int dt' \int \frac{d^3 {\bf q}}{(2 \pi)^3}
\left\{\lambda_\psi^2
{\rm Im} \left[ G_\psi^{++} ({\bf q},t-t') \right]^2 
+ 4 g^4  {\rm Im} \left[ G_\phi^{++} ({\bf q},t-t') \right]^2 
\right\}\;,
\nonumber \\ 
{}F_3 (t,t') &=&-2 g^2\int dt' \int \frac{d^3 {\bf q}}{(2 \pi)^3}
\left\{\lambda_\phi
{\rm Im} \left[ G_\phi^{++} ({\bf q},t-t') \right]^2 
+ \lambda_\psi {\rm Im} \left[ G_\psi^{++} ({\bf q},t-t') \right]^2 
\right\}\;,
\nonumber \\
{}F_4 (t,t') &=&-8 g^4 \int dt'  \int \frac{d^3 {\bf q}}{(2 \pi)^3}
{\rm Im} \left[ G_\phi^{++} ({\bf q},t-t') G_\psi^{++} ({\bf q},t-t')\right]
\;,
\label{Fs}
\end{eqnarray}

\noindent
and $\bar{\lambda_\phi}$, $\bar{\lambda_\psi}$ and $\bar{g}^2$ 
(the one-loop effective coupling constants) in 
Eqs. (\ref{eqphi2_1}) and (\ref{eqpsi2_1}) are given by

\begin{eqnarray}
\bar{\lambda_\phi} &=& \lambda_\phi +
\int dt' \int \frac{d^3 {\bf q}}{(2 \pi)^3}
\left\{\frac{\lambda_\phi^2}{2}
{\rm Im} \left[ G_\phi^{++} ({\bf q},t-t') \right]^2 
+ 2 g^4  {\rm Im} \left[ G_\psi^{++} ({\bf q},t-t') \right]^2 
\right\}\Big|_{t'=t}\;,
\nonumber \\
\bar{\lambda_\psi} &=& \lambda_\psi +
\int dt' \int \frac{d^3 {\bf q}}{(2 \pi)^3}
\left\{\frac{\lambda_\psi^2}{2}
{\rm Im} \left[ G_\psi^{++} ({\bf q},t-t') \right]^2 
+ 2 g^4  {\rm Im} \left[ G_\phi^{++} ({\bf q},t-t') \right]^2 
\right\}\Big|_{t'=t}\;,
\nonumber \\
\bar{g}^2 &=& g^2 + g^2\int dt' \int \frac{d^3 {\bf q}}{(2 \pi)^3}
\left\{\lambda_\phi
{\rm Im} \left[ G_\phi^{++} ({\bf q},t-t') \right]^2 
+ \lambda_\psi {\rm Im} \left[ G_\psi^{++} ({\bf q},t-t') \right]^2 
\right. \nonumber \\
&+& \left. 8 g^2 
{\rm Im} \left[ G_\phi^{++} ({\bf q},t-t') G_\psi^{++} ({\bf q},t-t')\right]
\right\}\Big|_{t'=t}\;,
\label{ctes}
\end{eqnarray}

\section{Coupling to Bath Degrees of Freedom and an Approximate
{}Form for the EOMs}

The two coupled nonlocal EOMs, Eqs. (\ref{eqphi2_1}) and
(\ref{eqpsi2_1}) (or equivalently, Eqs. (\ref{eqphi2}) and
(\ref{eqpsi2})) are too complicated to directly
numerically work with them. 
The main difficulty in handling these equations comes from
the dissipative like terms in Eqs. (\ref{eqphi2_1})
and (\ref{eqpsi2_1}), which have the non-Markovian kernels
shown in  Eq. (\ref{Fs}).  
If we suppose that there is a Markovian limit for those kernels,
then by using Eqs. (\ref{G of k}) and (\ref{G><}) in
Eq. (\ref{Fs}), we find that at $T=0$ the kernels diverge logarithmic
(a result also found in Ref. \cite{boya} for the single scalar field
case). However, in the high temperature limit ($\Gamma/\omega \ll 1$ and
$\Gamma/T \ll 1$) it has been argued in
Ref. \cite{GR} that a Markovian approximation  exists and a finite result
for the dissipation coefficients can be found. Such  an approximation
for the kernels given in Eq. (\ref{Fs}), we can also find here for our 
specific problem, in which case we may then, in principle, write
for the dissipation terms in  Eqs. (\ref{eqphi2_1}) and
(\ref{eqpsi2_1}) the approximate expressions

\begin{eqnarray}
\phi_c (t) \int_{-\infty}^t dt' 
\phi_c (t') \dot{\phi}_c (t') F_1 (t,t')
&\sim& \phi_c^2 (t) \dot{\phi}_c (t) \eta_1 \;,\nonumber \\
\psi_c (t) \int_{-\infty}^t dt' 
\psi_c (t') \dot{\psi}_c (t') F_2 (t,t')
&\sim& \psi_c^2 (t) \dot{\psi}_c (t) \eta_2 \;,\nonumber \\
\phi_c (t) \int_{-\infty}^t dt' 
\psi_c (t') \dot{\psi}_c (t') F_3 (t,t')
&\sim& \phi_c (t) \psi_c (t) \dot{\psi}_c (t) \eta_3 \;,\nonumber \\
\psi_c (t) \int_{-\infty}^t dt' 
\phi_c (t') \dot{\phi}_c (t') F_3 (t,t')
&\sim& \phi_c (t) \psi_c (t) \dot{\phi}_c (t) \eta_3 \;,\nonumber \\
\psi_c (t)
\int_{-\infty}^{t} dt' \left[\phi_c (t') \dot{\psi}_c (t')+
\dot{\phi}_c (t') \psi_c (t')\right] F_4 (t,t') 
&\sim& \psi_c (t)\left[\phi_c (t) \dot{\psi}_c (t)+
\dot{\phi}_c (t) \psi_c (t)\right] \eta_4 \;,\nonumber \\ 
\phi_c (t)
\int_{-\infty}^{t} dt' \left[\phi_c (t') \dot{\psi}_c (t')+
\dot{\phi}_c (t') \psi_c (t')\right] F_4 (t,t') 
&\sim& \phi_c (t)\left[\phi_c (t) \dot{\psi}_c (t)+
\dot{\phi}_c (t) \psi_c (t)\right] \eta_4 \;,
\label{markov}
\end{eqnarray}

\noindent
where $\eta_1, \eta_2,\eta_3$ and $\eta_4$ in the above expressions
are given by (using Eqs. (\ref{G of k}) and (\ref{G><}) in
Eqs. (\ref{eqphi2_1}) and
(\ref{eqpsi2_1}), and in the high temperature approximation
$\Gamma_\phi/T,\Gamma_\psi/T \ll 1$, with 
$\Gamma_\phi/\omega_\phi,\Gamma_\psi/\omega_\psi, \ll 1$ 
and
$\bar{m}^2_\phi \neq
\bar{m}^2_\psi$)

\begin{eqnarray}
\eta_1 &\sim& \frac{\lambda_\phi^2}{8} \beta \int \frac{d^3 q}{(2 \pi)^3}
\frac{n_\phi (1 + n_\phi)}{\omega_\phi^2  \Gamma_\phi }
+ \frac{g^4}{2} \beta \int \frac{d^3 q}{(2 \pi)^3}
\frac{n_{\psi} (1 + n_{\psi})}{\omega_{\psi}^2 
\Gamma_{\psi} } +
{\cal O}\left(\lambda_\phi^2 \frac{\Gamma_\phi}{\omega_\phi}, \; 
g^4 \frac{\Gamma_{\psi}}{\omega_{\psi}} \right) \;,\nonumber \\
\eta_2 &\sim& \frac{\lambda_\psi^2}{8} \beta \int \frac{d^3 q}{(2 \pi)^3}
\frac{n_\psi (1 + n_\psi)}{\omega_\psi^2 \Gamma_\psi }
+ \frac{g^4}{2} \beta \int \frac{d^3 q}{(2 \pi)^3}
\frac{n_{\phi} (1 + n_{\phi})}{\omega_{\phi}^2 
\Gamma_{\phi} } +
{\cal O}\left(\lambda_\psi^2 \frac{\Gamma_\psi}{\omega_\psi}, \; 
g^4 \frac{\Gamma_{\phi}}{\omega_{\phi}} \right) \;,\nonumber \\
\eta_3 &\sim& \beta\frac{\lambda_\phi g^2}{4} \int \frac{d^3p}{(2\pi)^3}
\frac{n_\phi (1 + n_\phi)}{\omega_\phi^2 \Gamma_\phi } 
+
\beta\frac{\lambda_\psi g^2}{4} \int \frac{d^3p}{(2\pi)^3}
\frac{n_\psi (1 + n_\psi )}{\omega_\psi^2 \Gamma_\psi } +
{\cal O}\left(\lambda_\phi g^2 \frac{\Gamma_\phi}{\omega_\phi}, \; 
\lambda_\psi g^2 \frac{\Gamma_{\psi}}{\omega_{\psi}} \right)\;,\nonumber\\
\eta_4  &\sim& 
8 g^4 \beta \int \frac{d^3 q}{(2 \pi)^3} \frac{1}{(\omega_\phi^2 - 
\omega_\psi^2)^2} \left[
n_{\phi} (1 + n_{\phi}) 
\Gamma_{\phi} +n_{\psi} (1 + n_{\psi}) 
\Gamma_{\psi} \right] +
{\cal O}\left(g^4 \frac{\Gamma_\phi^2}{\omega_\phi^2}, \; 
g^4 \frac{\Gamma_{\psi}^2}{\omega_{\psi}^2} \right) \;.
\label{disscoefs}
\end{eqnarray}

Similar time non-localities as the ones
appearing in Eqs. (\ref{eqphi2}) and (\ref{eqpsi2}) have 
also been dealt with in
Refs. \cite{GR,BGR} by using an adiabatic (or sudden) approximation for the
fields. The dissipation coefficients obtained by that approximation are
just the same as the ones given in (\ref{disscoefs}).
As shown in Ref. \cite{BGR} the adiabatic approximation
for the nonlocal kernels is a consistent approximation in the
case the fields are in an overdamped regime. The strong field
dissipation responsible for the overdamped regime can be attained by
coupling both $\Phi$ and $\Psi$ fields to a large set of other fields 
making the bath in which $\Phi$ and $\Psi$ evolve,  which then enlarges 
the number of field decay channels
available for both fields. This idea was used in Ref. \cite{BGR2} for the 
construction of an alternative inflationary model. 

\subsection{The Validity of the One-Loop and Markovian Approximations}

The consideration of the coupling of $\Phi$ and $\Psi$ 
to a large set of bath field degrees of freedom is particularly 
relevant to the validity of the Markovian approximation taken 
for the dissipative kernels and in order to make a clear 
assessment of the regime of validity not only of this approximation
but also for the validity of the one-loop approximation we used
to derive the EOMs. The study of the validity of these approximations
are clearly important to our study which is the study of 
chaotic behavior in the dynamics of our system of equations, and by 
making clear that chaos in our dynamical system is not just
an artifact of the approximations taken.
In order to address these important issues, let us consider  
the coupling of both $\Phi$ and $\Psi$ to a set of
$N$ (scalar) fields $\chi$ making the bath, which are coupled in the 
following way:

\begin{equation}
f_\phi^2 \sum_{i=1}^N \Phi^2 \chi_i^2 + f_\psi^2 \sum_{i=1}^N
\Psi^2 \chi_i^2 \;.
\label{bath}
\end{equation}

We next study how our main quantities scale with the various 
coupling constants and $N$ in the large N limit\footnote{R.O.R.
deeply thanks S. Jeon for correspondence regarding the large N contributions
to dissipation and the consistency of the adiabatic approximation in
the large N limit.}. This analysis will allow us to make a qualitative
and clear assessment of our main approximations.  

Let us first consider that the various coupling constants we have
scale with $N$ in the following way:

\begin{equation}
\lambda_\phi \sim \lambda_\psi \sim g^2 \sim \frac{\alpha}{\sqrt{N}} \;,
\label{model ctes}
\end{equation}

\noindent
and

\begin{equation}
f_\phi^2 \sim f_\psi^2 \sim \frac{\alpha^2}{N} \;,
\label{bath ctes}
\end{equation}

\noindent
with $\alpha \ll 1$. The choice for the coupling constants taken above 
is consistent
in the sense that higher order quantum corrections to quantities like
the field masses and coupling constants will not blow up  in the
large N limit. {}For instance, as $N \to \infty$ the effective masses
$\bar{m}^2_\phi$ and $\bar{m}^2_\psi$ scale as (in the high temperature
limit)

\begin{eqnarray}
\bar{m}_\phi^2 &=& m_\phi^2 + {\cal O}(\lambda_\phi T^2) +
{\cal O}(g^2 T^2) + {\cal O}(N f_\phi^2 T^2) \simeq m_\phi^2 +
{\cal O}(\alpha^2 T^2)\;,
\nonumber \\
\bar{m}_\psi^2 &=& m_\psi^2 + {\cal O}(\lambda_\psi T^2) +
{\cal O}(g^2 T^2) + {\cal O}(N f_\psi^2 T^2) \simeq m_\psi^2 +
{\cal O}(\alpha^2 T^2)\;,
\label{masses largeN}
\end{eqnarray}

\noindent
and the effective coupling constants go like

\begin{eqnarray}
\bar{\lambda}_\phi &=& 
\lambda_\phi + {\cal O}(\lambda_\phi^2) + {\cal O}(g^4) + 
{\cal O}(N f_\phi^4) \simeq
\lambda_\phi + {\cal O}\left(\frac{\alpha^2}{N}\right)\;,
\nonumber \\
\bar{\lambda}_\psi &=& 
\lambda_\psi + {\cal O}(\lambda_\psi^2) + {\cal O}(g^4) + 
{\cal O}(N f_\psi^4) \simeq
\lambda_\psi + {\cal O}\left(\frac{\alpha^2}{N}\right)\;,
\nonumber \\
\bar{g}^2 &=& g^2 + {\cal O}(\lambda_\phi g^2) + {\cal O}(\lambda_\psi g^2) 
+ {\cal O}(g^4) + 
{\cal O}(N f_\phi^2 f_\psi^2) \simeq g^2 +
{\cal O}\left(\frac{\alpha^2}{N}
\right)\;.
\label{ctes largeN}
\end{eqnarray}

We next determine the scaling of the fields decay widths
$\Gamma_\phi$, $\Gamma_\psi$ and for those of the bath,
$\Gamma_{\chi_i}$, which determines
the time-scale for collisions for the system in interaction with
the (thermal) bath. We begin by noticing that the imaginary part of the field 
self-energies contributing to the decay widths are typically dominated
by momenta $|{\bf q}| \sim T$ \cite{jeon,BGR}. We can then find for
the decay widths the relations:

\begin{eqnarray}
\Gamma_\phi &=& {\cal O} (\lambda_\phi^2 T) +{\cal O} (g^4 T) +
{\cal O} (N f_\phi^4 T) \simeq {\cal O} \left(\frac{\alpha^2}{N}T\right)
\;,\nonumber \\
\Gamma_\psi &=& {\cal O} (\lambda_\psi^2 T) +{\cal O} (g^4 T) +
{\cal O} (N f_\psi^4 T) \simeq {\cal O} \left(\frac{\alpha^2}{N}T\right)
\;,\nonumber \\
\Gamma_{\chi_i} &=& {\cal O} (f_\phi^4 T) +
{\cal O} (f_\psi^4 T) \simeq {\cal O} \left(\frac{\alpha^4}{N^2}T\right)
\;.
\label{decay largeN}
\end{eqnarray}

\noindent
Using these in the expressions for the dissipative kernels,
Eq. (\ref{Fs}), we can then show that the 
dissipation coefficients $\eta_1,\ldots,\eta_4$, appearing
in Eq. (\ref{markov}), have magnitude given by

\begin{eqnarray}
\eta_1 &\simeq&  {\cal O} \left(\frac{\lambda_\phi^2}{\Gamma_\phi}\right)
+ {\cal O} \left(\frac{g^4}{\Gamma_\psi}\right)
+  {\cal O} \left(N\frac{f_\phi^4}{\Gamma_{\chi_i}} \right)
\sim {\cal O} \left( \frac{N}{T} \right) \;,
\nonumber \\
\eta_2 &\simeq&  {\cal O} \left(\frac{\lambda_\psi^2}{\Gamma_\psi}\right)
+ {\cal O} \left(\frac{g^4}{\Gamma_\phi}\right)
+  {\cal O} \left(N\frac{f_\psi^4}{\Gamma_{\chi_i}} \right)
\sim {\cal O} \left( \frac{N}{T} \right) \;,
\nonumber \\
\eta_3 &\simeq&  {\cal O} \left(\frac{g^2\lambda_\phi}{\Gamma_\phi}\right)
+ {\cal O} \left(\frac{g^2 \lambda_\psi}{\Gamma_\psi}\right)
+  {\cal O} \left(N\frac{f_\phi^2 f_\psi^2}{\Gamma_{\chi_i}} \right)
\sim {\cal O} \left( \frac{N}{T} \right) \;,
\nonumber \\
\eta_4 &\simeq&  {\cal O} \left(g^4 \Gamma_\phi+g^4 \Gamma_\psi \right)
\sim {\cal O} \left( \frac{\alpha^4}{N^2 T} \right)\;.
\label{disscoef largeN}
\end{eqnarray}

\noindent
Therefore, the above estimates show that, for $\alpha^2 \ll 1$ and
in the large N limit, the dissipation kernel $F_4$ and its related
dissipation coefficient in the Markov approximation are negligible
compared with the others one. The behavior of the remaining coefficients
reproduce the typical behavior found previously for the dissipation
coefficients in Refs. \cite{GR} and \cite{BGR}. Additionally, we can 
also find that higher loop contributions to dissipation
(for instance the ones coming from two-loop diagrams and higher)
are all at most of order ${\cal O} (\lambda_\phi^2 \Gamma_\phi)$,
${\cal O} (\lambda_\psi^2 \Gamma_\psi)$, 
${\cal O} (g^4 \Gamma_\phi+g^4 \Gamma_\psi)$ or
${\cal O} (N f^4  \Gamma_\chi)$, which are all subleading in the 
perturbative regime. All these simple qualitative estimative allow
us then to assess the validity of both the Markov and one-loop approximations
used in this study.

\section{The Dynamical System and Chaotic Behavior}

{}From the results obtained in the last Section, we can then 
write Eqs. (\ref{eqphi2_1}) and
(\ref{eqpsi2_1}) in the local form shown below and more
suitable for the numerical analysis as:

\begin{eqnarray}
\ddot{\phi}_c  +  \bar{m}_\phi^2 \phi_c   +  
\frac{\bar{\lambda}_\phi}{6} 
\phi_c^3   +  \bar{g}^2 
\phi_c \psi_c^2  + 
\eta_1 \phi_c^2  \dot{\phi_c}  +  
\eta_3 \phi_c \psi_c \dot{\psi}_c  =  0 
\label{eqphi3}
\end{eqnarray}

\noindent
and
\begin{eqnarray}
\ddot{\psi}_c  +  \bar{m}_\psi^2 \psi_c   +  
\frac{\bar{\lambda}_\psi}{6} 
\psi_c^3   +  \bar{g}^2 
\psi_c \phi_c^2 + 
\eta_2 \psi_c^2  \dot{\psi_c}  +   
\eta_{\phi \psi} \phi_c \psi_c \dot{\phi}_c  =  0 ,
\label{eqpsi3}
\end{eqnarray}

\noindent 
where $\bar{\lambda}_{\phi} \sim \lambda_\phi,\bar{\lambda}_\psi
\sim \lambda_\psi$ and  $\bar{g}^2 \sim g^2$
denote the renormalized couplings
\footnote{The renormalization of
the coupling constants and masses can be made by  just the standard 
way, by introducing the appropriate counterterms of renormalization 
in our model Lagrangian \cite{GR,BGR}.}.
$\eta_1$, $\eta_2$ and
$\eta_3$ denote the dissipation coefficients and we have neglect
the term involving $\eta_4$ due to the estimates shown in
(\ref{disscoef largeN}). Although these coefficients
can be explicitly evaluated, however, we here
refrain ourselves from an explicit evaluation of these terms, which, as 
shown in the previous Section,
depend on the various parameters of the model, temperature and on the
number $N$ of other field degrees making the environment that $\Phi$ and
$\Psi$ may be coupled to. In fact, the magnitude of the dissipation
terms may be controlled by these additional field couplings, as shown in
the simple estimates obtained in the previous Section. 
Therefore, for the sake of simplicity we just take $\eta_1$,
$\eta_2$ and $\eta_3$ as additional free constant
parameters.

Next, let us define the following constants

\begin{eqnarray}
a^2 &=& \frac{\bar{m}_\phi^2}{6 |\bar{m}_\psi^2|}\;,\nonumber \\
 G^2 &=&
\frac{\bar{g}^2}{a^2 \bar{\lambda}_\psi}\;,\nonumber \\
\lambda_x &=&
\frac{\bar{\lambda}_\phi}{a^4 \bar{\lambda}_\psi}
\;,
\label{constants}
\end{eqnarray}

\noindent
in term of which we can define the dimensionless
variables:

\begin{eqnarray}
x &=& \sqrt{\bar{\lambda}_\psi} \frac{a^2}{\bar{m}_\phi} \, \phi_c
\;,\nonumber \\
z &=& \frac{a^2}{\bar{m}_\phi \langle \Psi \rangle_v}\, \dot{\phi}_c\;,
\nonumber \\ 
y &=& \frac{1}{\langle \Psi \rangle_v} \,\psi_c \;, 
\nonumber \\
w &=& \frac{1}{\sqrt{6 |\bar{m}_\psi^2|}
\langle \Psi \rangle_v}\, \dot{\psi}_c \;,
\label{variables}
\end{eqnarray}

\noindent
and by also rescaling time and
dissipation coefficients as $t' = \sqrt{6 |\bar{m}_\psi^2|} \, t$,
$\eta_1 = \eta_x \, a^4 \sqrt{\bar{\lambda}_\psi} / \langle \Psi 
\rangle_v$, $\eta_2 = \eta_y \, \sqrt{\bar{\lambda}_\psi} / 
\langle \Psi \rangle_v$ and $\eta_3 = \eta_{xy} \, a^2 
\sqrt{\bar{\lambda}_\psi} /
\langle \Psi \rangle_v$, respectively, we can then write Eqs.
(\ref{eqphi3}) and (\ref{eqpsi3}) in terms of the following
dimensionless first-order differential system of equations:

\begin{eqnarray}
\dot{x} &=& z 
\nonumber \\
\dot{z} &=& -a^2 \left(x + \frac{\lambda_x}{6} x^3 + G^2 x y^2 + 
\eta_x x^2 z
+ \eta_{xy} x y w \right) \nonumber \\
\dot{y} &=& w \nonumber \\
\dot{w} &=& \frac{y}{6} -\frac{y^3}{6} - G^2 x^2 y - \eta_y y^2 w
- \eta_{xy} x y z 
\;. 
\label{sys}
\end{eqnarray}

{}For convenience we also choose parameters such that $a^2 = 1$, $G^2 =
1$ and $\lambda_x=1$, which means we consider $\bar{m}_\phi^2 = 6
|\bar{m}_\psi^2|$ and $\bar{g}^2 = \bar{\lambda}_\phi =
\bar{\lambda}_\psi$, which is a choice consistent with the
considerations used in the previous Section,
Eq. (\ref{model ctes}). We also take as base values for the
(dimensionless) dissipation coefficients the values $\{ \eta
\}=(\eta_x,\eta_y,\eta_{xy})=(1/120,1/240, 1/200)$\footnote{
This is consistent with the general expressions shown for
the dissipation kernels ${}F_1,{}F_2$ and ${}F_3$, which for
$\lambda_\phi = \lambda_\psi = g^2$, satisfy the simple relation
between then: $2({}F_1 + {}F_2) = 5 {}F_3$.},
for which the
dynamics displayed by (\ref{sys}) happen in the weakly damped regime. 
We also make the special consideration that the temperature of the
system, which is kept fixed at a value $T$, is large enough such that
the high temperature approximation involved in the Markov limit
for the dissipation kernels is valid, but smaller than the critical 
temperature,
$T_{c_\psi}$, 
for phase transition in the $\Psi$ field direction, 
$T_{c_\psi} > T \gg \Gamma_\phi,\Gamma_\psi, \bar{m}_\phi,
\bar{m}_\psi$, where 
$T_{c_\psi}^2 \simeq  12 |m_\psi^2|/(g^2 + \lambda_\psi/2)$.

We then numerically solve the dynamical system with initial conditions
taken such that at $t=0$ the potential in the Lagrangian density
(\ref{Lpp}) is symmetrical in both field directions. The system is then
evolved in time till the symmetry breaking in the $\Psi$-field direction
occurs. We then look for chaotic regimes as $\eta_\phi$, $\eta_\psi$ and
$\eta_{\phi\psi}$ are changed.
Our choice for initial conditions to numerically solving the
(dissipative) dynamical system (\ref{sys}) is as follows. At the
initial time we consider 

\begin{equation}
(\phi_c, \dot{\phi}_c, \psi_c,
\dot{\psi}_c)|_{t=0} = (4 \Phi_{\rm cr}, 0,0,0)\;.
\label{IC}
\end{equation}

\noindent
A typical result for the evolution of the fields in time is
shown in {}Fig. 1, which already shows a  highly chaotic dynamics
prior to symmetry breaking.

{}Following the method of Box-Counting \cite{ott,levin,levin2,fractal},
around the initial
condition (\ref{IC}) it is then considered a box in phase space 
(for the dimensionless
variables) of size $10^{-5}$, inside which a large number of random
points are taken (a total of $200.000$ random points were used in each
run). All initial conditions are then numerically evolved by using an
eighth-order Runge-Kutta integration method and the fractal dimension is
obtained by statistically studying the outcome of each initial condition
at each run of the large set of points. Special care is taken to keep
the statistical error in the results always below $\sim 1 \%$. The
results obtained are shown in Table I for different values of
dissipation coefficients. In this table we also show the uncertainty 
exponent $\epsilon$ (see Ref.
\cite{fractal}), which gives a measure of how chaotic is the system.
Qualitatively speaking we can say that the closest is $\epsilon$ of
zero, the more chaotic is the system. On the contrary, the closest is
$\epsilon$ of unit, the less chaotic is the system. We see clearly the
effect of dissipation on the nonlinearities of the system. It
can change fast from a chaotic to an integrable regime with a relatively
small increase of the dissipation. Larger dissipations tend also to
destroy fast the chaotic attractors. In {}Fig. 2 we show an example of
the structure of the chaotic attractors in the $\psi_c,\dot{\psi}_c$
plane ($y,w$ plane).

\section{Conclusions}

In this paper we have studied the dynamical system made by the 
leading order effective equations of motion for a model of
two coupled scalar fields.  
The chaotic behavior for (ensemble averaged) field trajectories
has been demonstrated and we have also shown that
in the overdamped regime
chaos gets completely suppressed. Chaotic motion can only develop in the
underdamped or very weakly damped regime, in which case enough energy
can be exchanged fast enough from one field to the other, making both
$\phi_c$ and $\psi_c$ to fluctuate with large enough amplitudes, leading
to a highly nonlinear behavior that then precludes the chaotic motion of
the system.

It seems also that we can indirectly associate the chaoticity in the
system with the equilibration rates of the fields, in close analogy with
the one found between the Lyapunov exponent and the thermalization rate
in perturbative thermal gauge theory \cite{chaos-thermal}. We
note from the results obtained here and from the numerical simulations
we performed, that the smaller is the fractal dimension (or the larger
is $\epsilon$) the fastest the fields equilibrate to their asymptotic
states by loosing their energies to radiation, which will then eventually
thermalize. A clear assessment to this interesting point, however,
must be carefully studied from the complete nonlocal EOMs, which then
restores the time reversal invariance of the equations of motion.

{}Finally we would like to point out that the kind of model we have
studied here and its generalization to larger number of fields is
natural to be found in extensions of the standard model with large
scalar sectors. Physical implementations of the model can be found, for
example, in particle physics or condensed matter models displaying
multiple stages of phase transitions, in which case the dynamics we have
studied here would be likely to manifest between any of that stages and,
therefore, with consequences to the phenomenology of that models. 
The model studied here has also a strong motivation from inflationary
models (hybrid inflation) \cite{linde}. In special, in this context,
a model Lagrangian of a similar form of the one we studied here
has been studied in Ref. \cite{maeda}, showing the possibility
of chaotic behavior during the final stages of inflation.
However, the authors in \cite{maeda} make use of the classical
equations of motion. It would be extremely interesting to assess the
effect of quantum effects (and particle productions) and consequently
dissipation also in that context, which is of fundamental 
relevance for the description of the process of reheating.
In the particular study performed here, our results could apply
instead to the description of the pre-inflationary stage,
with possible contributions to the discussion of the 
fine-tuning problem of the initial field configuration in hybrid
inflation \cite{tetradis}. 
These
problems are currently being examined by us and more results and details
will be reported elsewhere.

\bigskip
\acknowledgements

R.O.R. is partially supported by CNPq and F. A. R. N. was
supported by a M.Sc. scholarship from CAPES.

\begin{figure}[c]

\epsfysize=10cm 
{\centerline{\epsfbox{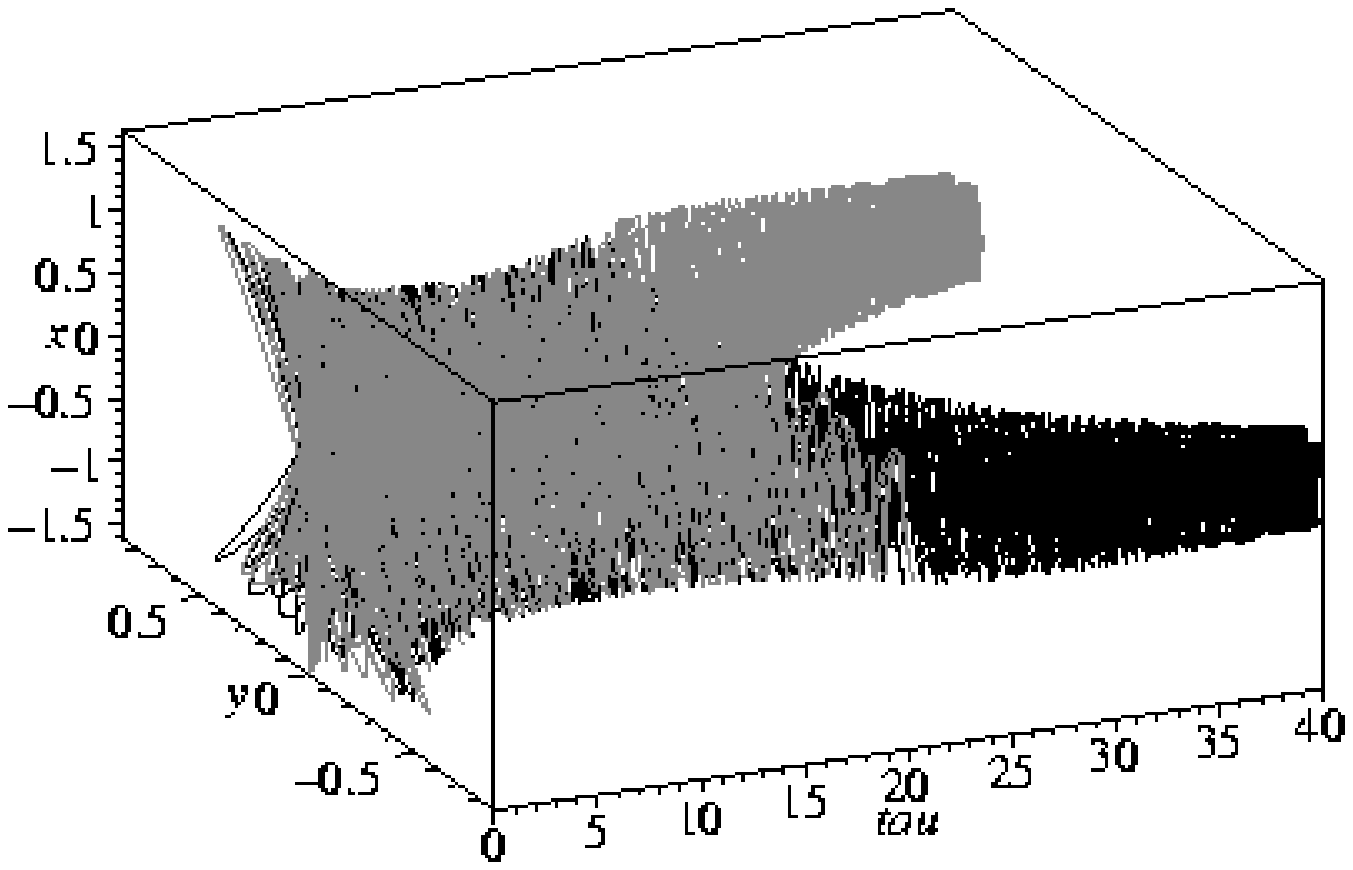}}}

\caption{An example of highly chaotic dynamics displayed by (\ref{sys}) 
for two initial conditions given (in the dimensionless variables)
by $(x,z,y,w)=(4/\sqrt{6},-10^{-5},0, \pm 10^{-5})$. Time is scaled as
$\tau= 10^2 \sqrt{6 |m_\psi^2|} \; t$. }

\end{figure}

\newpage

\begin{figure}[c]

\epsfysize=10cm 
{\centerline{\epsfbox{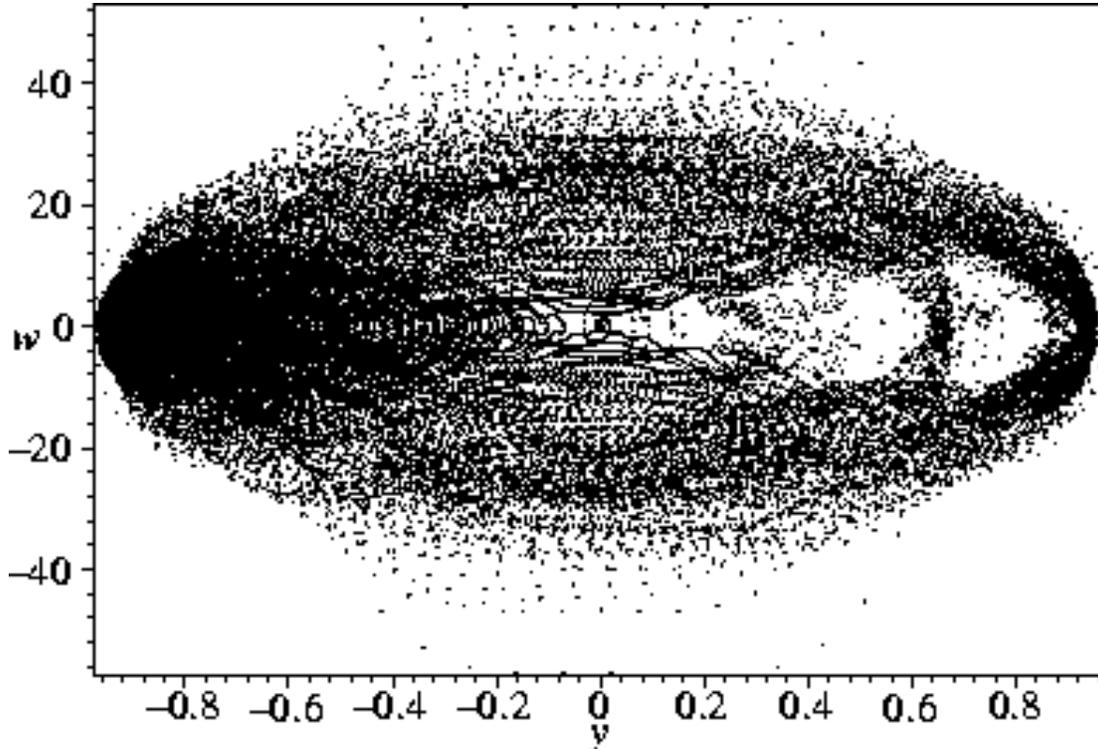}}}

\caption{The chaotic attractor for one realization of the initial condition 
$(x,z,y,w)=(4/\sqrt{6},-10^{-5},0, 10^{-5})$, for the first
dissipation case shown in Table I.}

\end{figure}

\begin{table} 
\caption{The fractal dimension $D_f$ and the uncertainty exponent 
$\epsilon$ for increasing dissipation coefficients ($D=4$ is the phase 
space dimension).}  
\begin{tabular}{ccc} 
$\{ \eta \} = (\eta_x,\eta_y,\eta_{xy})$ & $D_f$ & 
$\epsilon=D-D_f$ \\
\hline  
0.50 $\{\eta\}$    & 3.96   &   0.04  \\  
0.75 $\{\eta\}$    & 3.90   &   0.10  \\  
1.00 $\{\eta\}$    & 3.43   &   0.57  \\  
1.25 $\{\eta\}$    & 3.23   &   0.77  \\  
1.50 $\{\eta\}$    & 3.10   &   0.90  \\  
\end{tabular} 
\end{table}


\begin{references}

\bibitem{boya1}D. Boyanovsky and H. J. de Vega,
hep-ph/9909372, in the Proceedings of the IV Paris
Cosmology Colloquium (in press).

\bibitem{biro} T. S. Bir\'o, S. G. Matinyan and B. M\"uller, {\it
Chaos and Gauge Field Theory}, (World Scientific, Singapore,
1994). 

\bibitem{GR} M. Gleiser and R. O. Ramos, Phys. Rev. {\bf D50}, 2441
(1994);  M. Morikawa and M. Sasaki, Phys. Lett. {\bf 165B}, 59 (1985).

\bibitem{boya} D. Boyanovsky, H. J. de Vega, 
R. Holman, D. S-Lee and A. Singh, Phys. Rev. {\bf D51},
4419 (1995). 

\bibitem{greiner} C. Greiner and B. M\"uller, Phys. Rev. {\bf D55},
1026 (1997).

\bibitem{BGR} A. Berera, M. Gleiser and R. O. Ramos, Phys. Rev. {\bf D58},
123508 (1998). 

\bibitem{BGR2} A. Berera, M. Gleiser and R. O. Ramos, Phys. Rev. Lett.
{\bf 83}, 267 (1999). 

\bibitem{hu}E. Calzetta and B. L. Hu, Phys. Rev. {\bf D61}, 025012 (2000);
Phys. Rev. {\bf D40}, 656 (1989). 

\bibitem{ian}I. D. Lawrie, Phys. Rev. {\bf D60}, 063510 (1999);
J. Phys. {\bf A25}, 6493 (1992).

\bibitem{ott} E. Ott, {\sl Chaos in Dynamical Systems} (Cambridge
University Press, Cambridge 1993).

\bibitem{levin} N. J. Cornish and J. J. Levin, Phys.
Rev. {\bf D53}, 3022 (1996); Phys. Rev. {\bf D55}, 7489 (1997).

\bibitem{levin2} J. D. Barrow and J. Levin, Phys. Rev. Lett. 
{\bf 80},  656 (1998).

\bibitem{tadpole}G. Semenoff and N. Weiss, Phys. Rev. {\bf D31}, 699 (1985);
A. Ringwald, Ann. Phys. {\bf 177}, 129 (1987); Phys. Rev.
{\bf D36}, 2598 (1987).

\bibitem{bazeia}V. Latora and D. Bazeia,  Int. J. Mod. Phys. {\bf A14},
4967 (1999).

\bibitem{jeon}S. Jeon, Phys. Rev. {\bf D52}, 3591 (1995).

\bibitem{fractal}L. G. S. Duarte, 
L. A. C. P. da Mota, H. P. de Oliveira, R. O. Ramos and J. E. Skea,
Comp. Phys. Comm. {\bf 119}, 256 (1999).

\bibitem{chaos-thermal} 
U. Heinz, C. R. Hu, S. Leupold, S. G. Matinian and
B. M\"uller, Phys. Rev. {\bf D55}, 2464 (1997); 
T. S. Biro and M. H. Thoma, Phys. Rev. {\bf D54}, 3465 (1996);
T.S. Biro, C. Gong and B. M\"uller, Phys. Rev. {\bf D52},
1260 (1995). 

\bibitem{linde}A. D. Linde, Phys. Lett. {\bf B259}, 38 (1991);
Phys. Rev. {\bf D49}, 748 (1994).

\bibitem{maeda}R. Easther and K. Maeda, Class. Quant. Grav.
{\bf 16}, 1637 (1999).

\bibitem{tetradis}N. Tetradis, Phys. Rev. {\bf D57}, 5997 (1998);
C. Panagiotakopoulos and N. Tetradis,
Phys. Rev. {\bf D59}, 083502 (1999).


\end{references}
\end{document}